\pdfoutput=1

\documentclass[runningheads,a4paper]{llncs}

\usepackage{amssymb,amsmath}
\usepackage{graphicx}
\usepackage{xspace}     
\usepackage{color}
\usepackage{url}
\usepackage{multirow}
\usepackage{subfigure}
\usepackage{adjustbox}
\usepackage[table]{xcolor}

\urldef{\mailsc}\path|{csdwu, csrchang}@comp.polyu.edu.hk|

\newcommand{\vul}{FileCross\xspace}
\newcommand{\file}{\texttt{file://}\xspace}
\newcommand{\myfig}{Fig.\xspace}

\newcommand{\no}{n}
\newcommand{\yes}{\textcolor[rgb]{1.00,0.00,0.00}{y}}

\newcommand{\apJS}{A}
\newcommand{\apReport}{C}
\newcommand{\apNew}{B}
\newcommand{\apRes}{D}

\title{Analyzing Android Browser Apps for\\ \file Vulnerabilities}
\titlerunning{Analyzing Android Browser Apps for \file Vulnerabilities}

\author{Daoyuan Wu \and Rocky K. C. Chang}
\authorrunning{D. Wu and R.K.C. Chang}

\institute{Department of Computing, The Hong Kong Polytechnic University\\
\mailsc\\
}

\begin{document}

\mainmatter  

%
%

\toctitle{Analyzing Android Browser Apps for file:// Vulnerabilities}
\tocauthor{Daoyuan Wu and Rocky K. C. Chang}
\maketitle

\begin{abstract}



Securing browsers in mobile devices is very challenging, because these browser apps usually provide browsing services to other apps in the same device. A malicious app installed in a device can potentially obtain sensitive information through a browser app. In this paper, we identify four types of attacks in Android, collectively known as \vul, that exploits the vulnerable \file to obtain users' private files, such as cookies, bookmarks, and browsing histories.
We design an automated system to dynamically test 115 browser apps collected from Google Play and find that 64 of them are vulnerable to the attacks. Among them are the popular Firefox, Baidu and Maxthon browsers, and the more application-specific ones, including UC Browser HD for tablet users, Wikipedia Browser, and Kids Safe Browser. A detailed analysis of these browsers further shows that 26 browsers (23\%) expose their browsing interfaces unintentionally. In response to our reports, the developers concerned promptly patched their browsers by forbidding \file access to private file zones, disabling JavaScript execution in \file URLs, or even blocking external \file URLs. We employ the same system to validate the ten patches received from the developers and find one still failing to block the vulnerability.
\end{abstract}

\section{Introduction}
\label{sec:intro}

Using \file to browse local files is very common in desktop browsers. However, this file protocol mechanism, when applied to mobile platforms, could cause unexpected security risks. In modern smartphone systems, notably Android and iOS, each app's sensitive files are stored in their own system-provided private file zones, which cannot be accessed by other apps or users. Supporting \file without additional access control in mobile browsers, however, will break such security boundaries. This \file vulnerability is further aggravated in Android, because Android browsers usually accept external browsing requests
which, in the absence of any user interaction, can be issued by another (malicious) app. Unlike Android, these requests in iOS must be invoked by users' clicking.


Supporting external \file browsing requests (or termed as external \file URLs) is only a necessary condition for realizing actual attacks. In this paper, we show that combining with the capability of accessing private file zones through \file, JavaScript support, and other browsers' flaws (such as auto-file download), a malicious app in Android can launch four different types of attacks to steal a victim browser's private files (e.g., users' cookies, bookmarks, and browsing histories) or a victim website's private files (e.g., cookie or content). We refer to this class of attacks as \textit{\vul}, in which all attack vectors are delivered through the \file protocol between a browser app and an attack app. The attack app can automatically download a private file to the public SD card for exporting, steal a private file by compromising same-origin policy (SOP~\cite{SOP}) on the ``host'' level, steal the content of another website by compromising SOP on the protocol level (\file and \texttt{http(s)://}), and steal a private file by exploiting a SOP flaw in handling symbolic links.

Several isolated incidences on stealing browsers' private files were reported for Chrome and Firefox \cite{Download13,Symlink13,Firefox13}. However, as we will show in this paper, these attacks are just instances of the FileCross attacks. To characterize the prevalence and impact of the FileCross attacks, we develop a system based on dynamic analysis to automatically test over 100 browser apps in Android. The main approach is to mimic actual attacks and use them to test the browsers on real smartphones. This system determines whether a browser app is vulnerable to the four FileCross attacks. It also analyzes whether the app, before and after patching, supports \file, allows access to private file zones through \file, and supports JavaScript.

The main findings obtained from our analysis of 115 browser apps can be summarized below.
\begin{enumerate}
  \item More than half of the browsers tested are vulnerable to the \vul attacks. In particular, 50\% of the most popular browsers (e.g., Firefox, Baidu, and Maxthon) are also vulnerable. Similarly, many major browsers in different categories could leak out private information through the \vul attacks. Among the four different attacks, the three attacks that are based on compromising SOP affect 55\% of the browsers on Android 4.0, 4.3 or 4.4.

  \item The \file vulnerabilities are exploitable in all Android versions (including the latest 4.4), and even occur in different web engines. Specifically, our system identifies 46 browsers being vulnerable in 4.4 (across all four \vul attacks). This result contradicts the general belief that Chrome-based new system engine will no longer contain these flaws by default. We are also contacting Google Android security team to fix one common flaw at the engine level. Moreover, we detect three vulnerable browsers (Firefox, UC Browser HD and Sogou) out of 15 browsers that employ custom engines. 

  \item A further analysis reveals that 23\% of the browsers expose their browsing interfaces unintentionally. Had the developers realized the browser interfaces' exposure, one third of them would not have been vulnerable to the \vul attacks. Moreover, 65\% of the browsers accept external \file browsing requests, and 62\% even allow \file access to the private file zones. The latter is necessary for three \vul attacks. Moreover, 63\% support JavaScript execution in \file URLs which makes three \vul attacks possible.

  \item In response to our vulnerability reports, 19 developers followed up with our findings. We have so far received nine patches from them (and will receive more). An analysis of the patches shows that the patching methods include disabling the access to unrenderable private files, blocking external \file URLs, or disabling JavaScript execution in \file URLs. Most of them could effectively thwart the attacks. However, our system developed for testing browsers finds that one patch failed to block the vulnerability, because the patch missed a second attack entry.
\end{enumerate}

\section{The \file Vulnerabilities}
\label{sec:problem}


\subsection{The FileCross Attacks}
\begin{figure*}[ht!]
\vspace{-2ex}
\begin{center}
  \centerline{\includegraphics[width=1.06\textwidth]{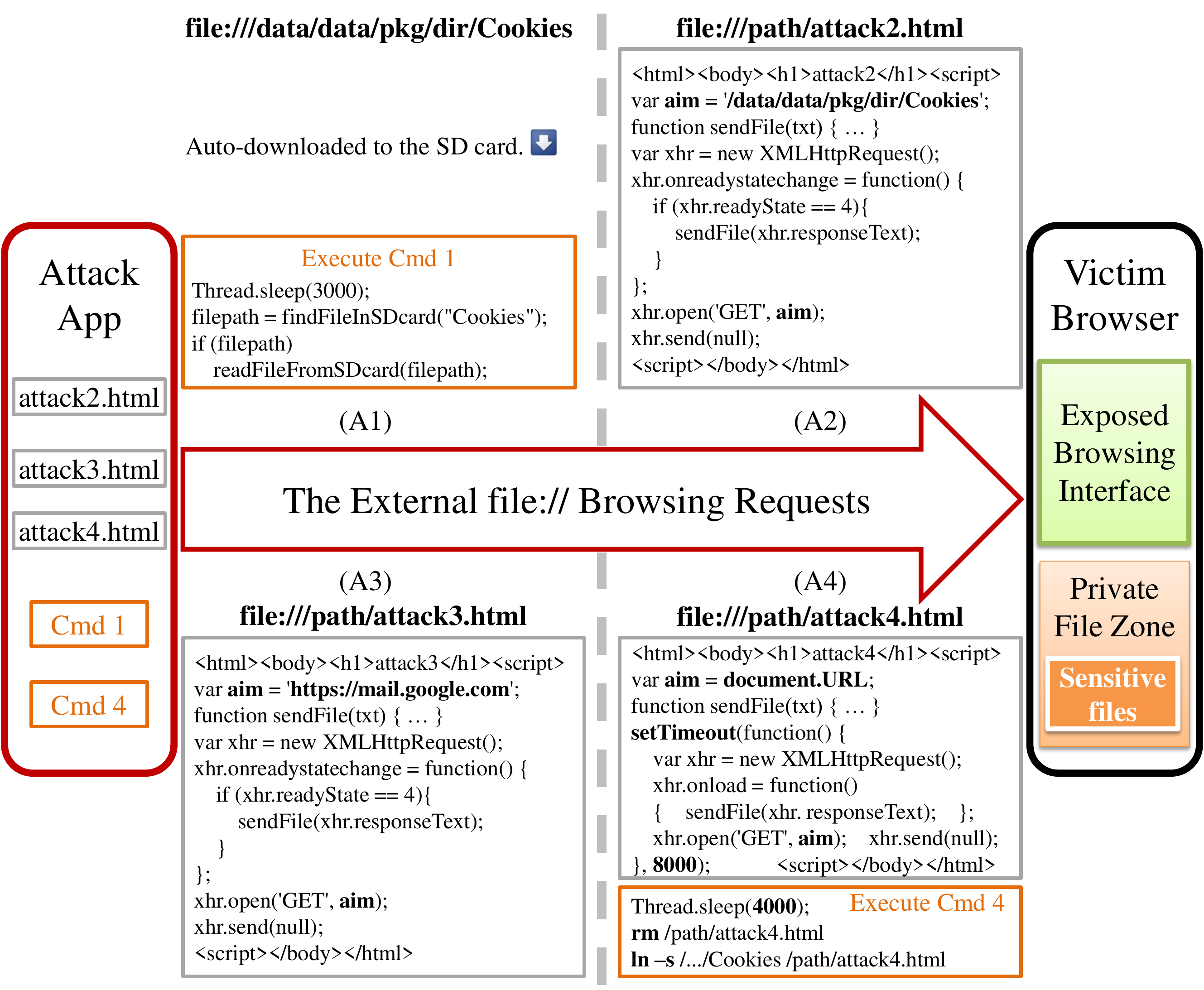}}
\end{center}
\vspace{-6ex}
\caption{Examples of four \vul attacks (A1 to A4).}
\vspace{-2ex}
\label{fig:attack}
\end{figure*}

We have discovered from our evaluation, which will be further elaborated in Section \ref{sec:atkcondition}, that 113 out of 115 browsers in Android expose their browsing interfaces, and 75 out of the 113 browsers support external browsing requests from other apps through \file. As illustrated in \myfig \ref{fig:attack}, an attack app can issue a ``malicious'' browsing request to a victim browser through the \file channel. The attack can steal sensitive files directly or indirectly from the victim browser's private file zone by having the URL in the browsing request point to a target sensitive file or a malicious HTML file, respectively.

The direct method exploits the fact that some browsers allow \file requests to access their private file zones. The indirect method, on the other hand, exploits the same-origin policy (SOP \cite{SOP}) flaws in handling \file requests, and it also requires the JavaScript support for executing the malicious HTML file. In our evaluation, 71 browser apps (out of the 75 that support \file) allow the requests received from \file to access their private file zones, and 72 permit JavaScript execution in \file URLs. Moreover, the indirect method can be used to steal sensitive files from websites.

\myfig \ref{fig:attack} shows examples of four \vul attack patterns. The first one uses the direct method, whereas the last three use the indirect method by compromising the SOP.
The first and fourth attacks are in fact first reported by an individual hacker.
We discovered the other two from the Android developer document.
We thus do not claim the discovery of these attacks as our main contribution.
But we are the first to identify them as a unified attack model (i.e., FileCross) and conduct automated testing to analyze their prevalence in Android browsers.
In addition, our system to be presented in Section \ref{sec:system} could be extended to detect new attack patterns. 
\begin{itemize}
  \item[Attack 1 (A1):] The \file URL points to a sensitive file (\texttt{Cookies} in the figure) in the victim browser's private file zone. Some browsers
    automatically download the requested file to the \texttt{Download} directory on a SD card. The attack app can use keyword search to find and read the target file from the SD card (see \texttt{Cmd 1}). The auto-download feature has been identified as a flaw responsible for a successful \vul attack against Chrome for Android \cite{Download13}.

  \item[Attack 2 (A2):] The \file URL points to a malicious HTML file \texttt{attack2.html}. The attacker prepares the HTML file for the browser to retrieve a sensitive file (\texttt{Cookies} in the figure) from its private file zone. Once the attack HTML file is loaded, an asynchronous request (e.g., via the XMLHttpRequest API~\cite{XMLHttpRequest}) is issued to retrieve the sensitive file (\texttt{xhr.responseText} in the figure). After this, \texttt{sendFile(txt)} is invoked to send the file to a remote server that can be accessed by the attacker. The fundamental problem enabling this attack is compromising SOP for \file requests (i.e., a local file should not be allowed to read contents of another file). Our evaluation shows that 63 browsers are vulnerable to this attack.

  \item[Attack 3 (A3):] The \file URL points to a malicious HTML file \texttt{attack3.html}. The attacker prepares the HTML file for the browser to retrieve sensitive content from a remote website (\texttt{mail.google.com} in the figure). Similar to the last attack, the content is retrieved by an asynchronous request and sent to a remote server via \texttt{sendFile(txt)}. The fundamental problem is again compromising SOP, but this time on the protocol level (\file and \texttt{https}). Our evaluation uncovers 56 vulnerable browsers. This attack can also steal cookies of a website, but the details are omitted here.

  \item[Attack 4 (A4):] The \file URL points to a malicious HTML file \texttt{attack4.html}. While the objective of this attack is the same as A2, it sets the target (in the \texttt{aim} JavaScript variable) as the current URL (i.e., \texttt{document.URL} in the figure), thus not violating SOP. However, the codes will not be executed until after 8000 ms. The attack app in the meantime removes \texttt{attack4.html} and builds a symbolic link for the removed file using the target sensitive file \texttt{Cookies}. Now when the time comes for the browser to execute the codes, it may load \texttt{Cookies} according to the link and return its contents to JavaScript. This flaw of loading a symbolic link to a file when the file cannot be found exists in modern browsers, including Chrome \cite{Symlink13} and Firefox \cite{Firefox13}. Our evaluation reveals 57 vulnerable browsers.
\end{itemize}

    The last three attacks exploit the flaws on enforcing SOP for external \file requests.
    For \textit{webkit}, Android's default web engine, the SDKs prior to 4.1 suffered from flawed SOP enforcement. Although the flaws in attacks A2 and A3 have been fixed by the default setting introduced to Android 4.1, the \file vulnerabilities still remain for two reasons. First, we notice that the two new APIs introduced in 4.1 
    still suffer from the SOP flaws. Therefore, developers may still use these vulnerable APIs, especially when they cannot find the security implications from Google's Developer Document. Second, developers must compile their apps using recent SDKs to block the vulnerabilities. Our evaluation, however, shows that over 30 browsers on Android 4.3 are still vulnerable, because the developers still used the old SDKs to compile their apps.


Starting from the latest Android 4.4, the system web engine is changed to Chrome's Blink engine.
A general belief is that Chrome-based engine will no longer contain these flaws by default (we even made this mistake earlier via preliminary manual testing, since file paths are changed in 4.4).
But surprisingly, our automated testing finds 46 browsers are still vulnerable in 4.4, across all four \vul attacks.
In particular, we notice Android 4.4 does not provide by-default patches for the SOP flaw (in A4), causing 40 browsers still exploitable in 4.4 by attack A4.
We are contacting Google Android security team to fix this common flaw at the engine level.
Moreover, similar to the Android 4.3 cases, apps compiled with old SDKs (i.e., below 4.1) cannot be protected by system-level defenses for attacks A2 and A3, even running on Android 4.4.
Additionally, the flaw in A1 is application specific.
In summary, mitigating the \vul flaws in all Android versions still require browser developers' careful implementations.
Therefore, our evaluation system is designed to test browser implementations but not specific web engines.

\subsection{Attack Conditions}
\label{sec:atkcondition}

Table \ref{tab:condition} summarizes the conditions required for launching the four \vul attacks. Exposing browsing interfaces and supporting \file are obviously necessary for all of them. Allowing \file access to private file zones is also necessary for major \vul attacks that aim at stealing browsers' private files.
In addition, attacks A2, A3, and A4 require JavaScript execution in \file URLs for constructing the corresponding exploits (as shown in \myfig \ref{fig:attack}).
Although it is always possible for some advanced attackers to invent non-JavaScript exploits for these three attacks, we believe this JavaScript condition is currently required and therefore include it into our \vul threat models. 

\begin{table}[ht!]
  \centering
  \vspace{-1ex}
  \caption{The required conditions for the four \vul attacks.}
  \vspace{-1ex}
  \label{tab:condition}
\scalebox{0.9}{
\begin{adjustbox}{center}
\begin{tabular}{ |c | c | c | c | c | c| }
\hline
\cellcolor{gray!15} & \multicolumn{5}{c|}{\cellcolor{gray!15}Required Attack Conditions} \tabularnewline
\cline{2-6}
\cellcolor{gray!15}Attack & Exposed & \ Support \ & \file access & JavaScript & \multirow{3}{*}{Major flaws} \tabularnewline
\cellcolor{gray!15}IDs & \ browsing \ & \file & to private & execution in & \tabularnewline
\cellcolor{gray!15} & interface & URLs & file zones & \file URLs &  \tabularnewline
\hline
A1 & $\surd$ & $\surd$ & $\surd$ &   & Auto-download file to SD card \tabularnewline
\hline
A2 & $\surd$ & $\surd$ & $\surd$ & $\surd$ & SOP bypass for two \file origins \tabularnewline
\hline
A3 & $\surd$ & $\surd$ &   & $\surd$ & SOP bypass for \file and \texttt{http(s)://} origins \tabularnewline
\hline
A4 & $\surd$ & $\surd$ & $\surd$ & $\surd$ & SOP bypass in handling symbolic links \tabularnewline
\hline
\end{tabular}
\end{adjustbox}
}
\vspace{-4ex}
\end{table}

\begin{figure}[ht!]
  \centering
  \vspace{-4ex}
\begin{adjustbox}{center}
  \subfigure[Intentionally or unintentionally exposed browsing interface and their related attributes.] {
	\label{fig:exposed}
    \includegraphics[height = 23ex]{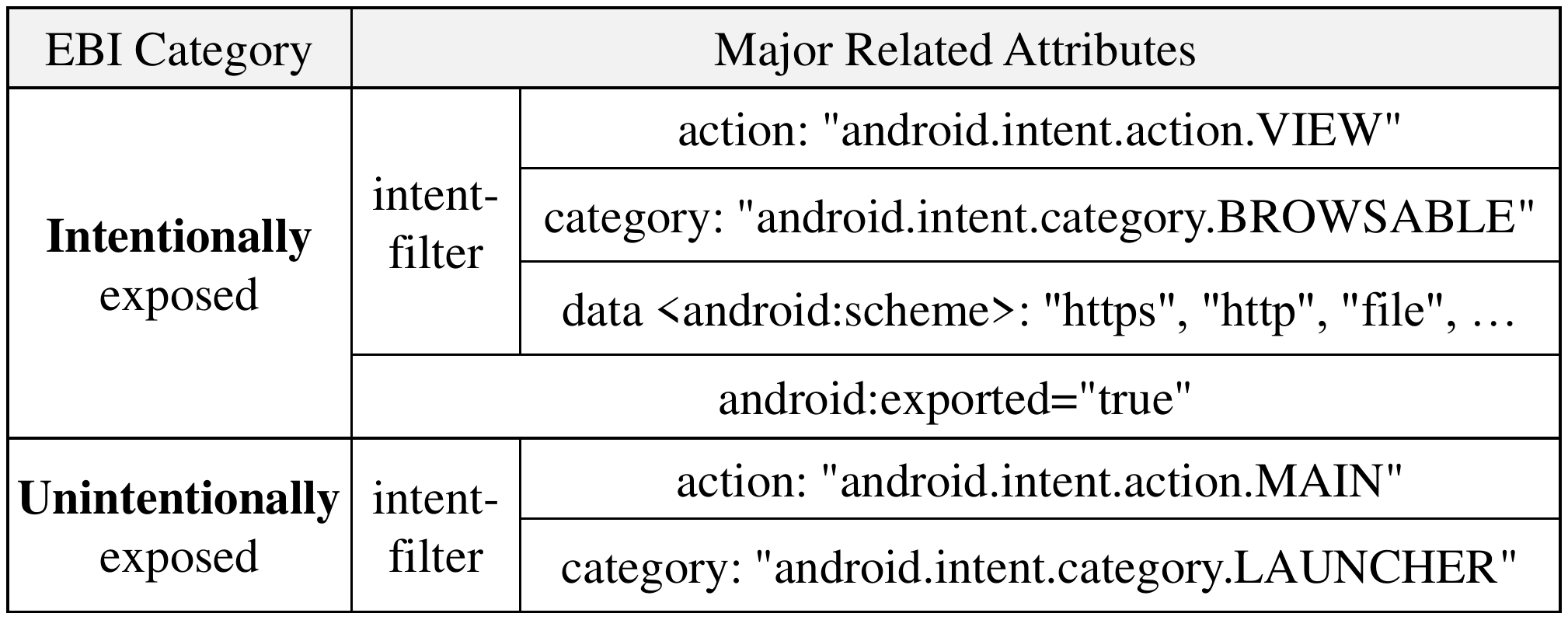}
  }
  \subfigure[The intentionally exposed browsing interface (\texttt{.ViewLink}) in Offline Browser (\texttt{it.nikodroid.offline}).] {
	\label{fig:offlineBrowser}
    \includegraphics[height = 23ex]{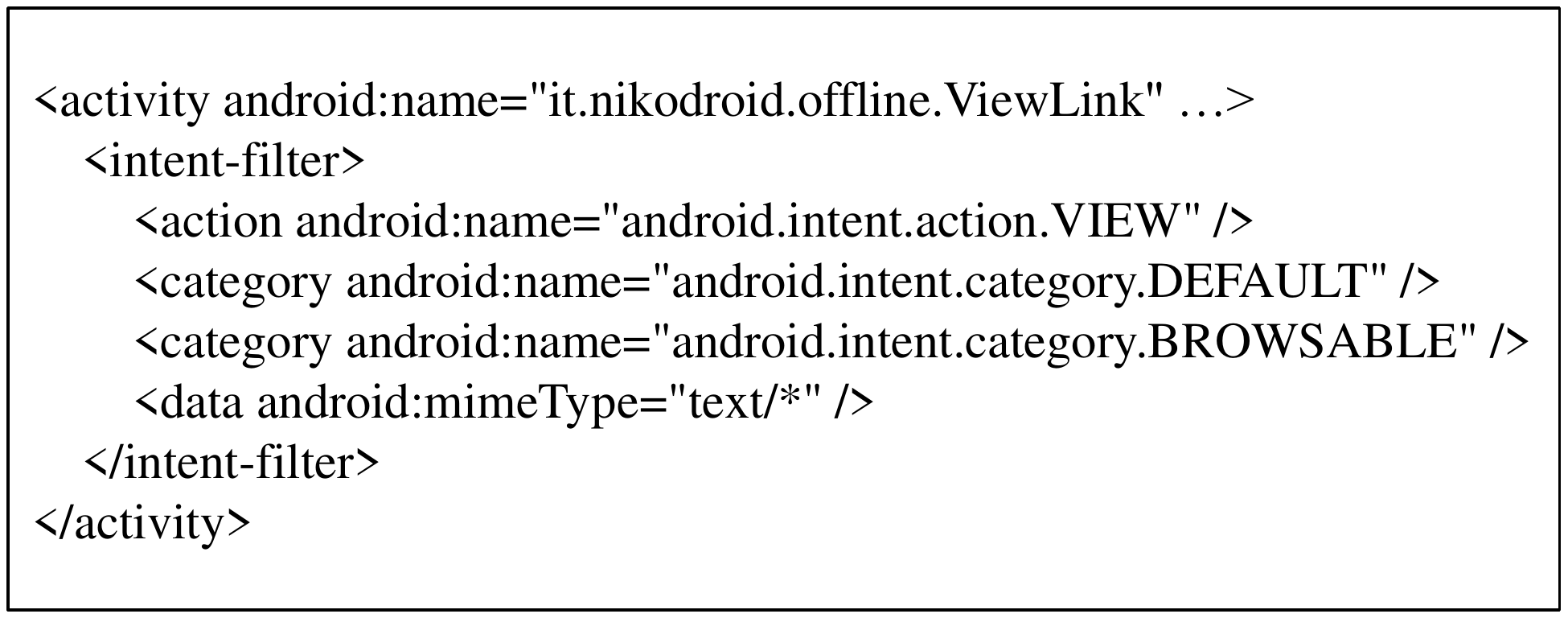}
  }
\end{adjustbox}
\vspace{-2ex}
\caption{A summary of EBI patterns and an EBI example.}
\vspace{-4ex}
\label{fig:exposeall}
\end{figure}

Before moving to the next section, it is instructive to understand how browsing interfaces are exposed. As mentioned above, 113 of our tested 115 browsers expose their browsing interfaces to other apps. By inspecting their manifest files, we further infer that some browsers expose their browsing interfaces \textit{unintentionally}, although most express \textit{explicit} intentions to accept external browsing requests.
We summarize these intentionally and unintentionally exposed patterns in \myfig \ref{fig:exposed}, and also give a simple Exposed Browsing Interface (EBI) example in \myfig \ref{fig:offlineBrowser}.
Our inference for intentional exposures is based on the presence of an Intent with the \texttt{action} of ``VIEW'' and the \texttt{category} of ``BROWSABLE,'' because this type of Intent is usually delivered to browsers \cite{BROWSABLE}.

The unintentionally exposed cases, in our understanding, are mainly caused by the Android's implicit Intent mechanism \cite{Intent}.
Specifically, Android requires each app to register an Intent filter with the \texttt{action} of ``MAIN'' and the \texttt{category} of ``LAUNCHER'' for the first user interface component, so that the app can be launched by the default launcher.
This behavior, however, will implicitly cause the corresponding component to be exposed to other apps.
It may happen for some browser developers to register their browsing interfaces with such Intent, thus exposing them as EBIs even without claiming to receive ``BROWSABLE'' intents.
Hence, these EBIs cannot be triggered by normal browsing requests. 
We thus believe they are unintentionally exposed by developers in terms of serving external browsing requests.
Due to the space limitation, we refer readers to Section 5.4 of \cite{ComDroid11} for a general discussion on such implicit intents.

\section{Automated Testing of Android Browsers}
\label{sec:system}


We design and implement a system for testing browsers for the \file vulnerabilities. In order to test all browser apps available in Android markets, our system can automatically test all of them without human intervention. Using the system, we could test over 100 Android browsers in less than four hours. Since our ultimate goal is to report vulnerable browsers to their developers for patching, it is not enough to just demonstrate that private files can be \textit{accessed} by invoking JavaScript's \texttt{alert(content)} function. Instead, our system mimicks the actual attacks to \textit{steal} victim browsers' private files and tests the browsers on actual smartphones. Besides detecting the vulnerabilities, the system also helps determine whether the external browsing interfaces are opened intentionally and analyze the patches obtained from the developers.


\subsection{The System Design}
\label{sec:overview}

\begin{figure*}[ht!]
\begin{center}
\vspace{-4ex}
\includegraphics[width=1.0\textwidth]{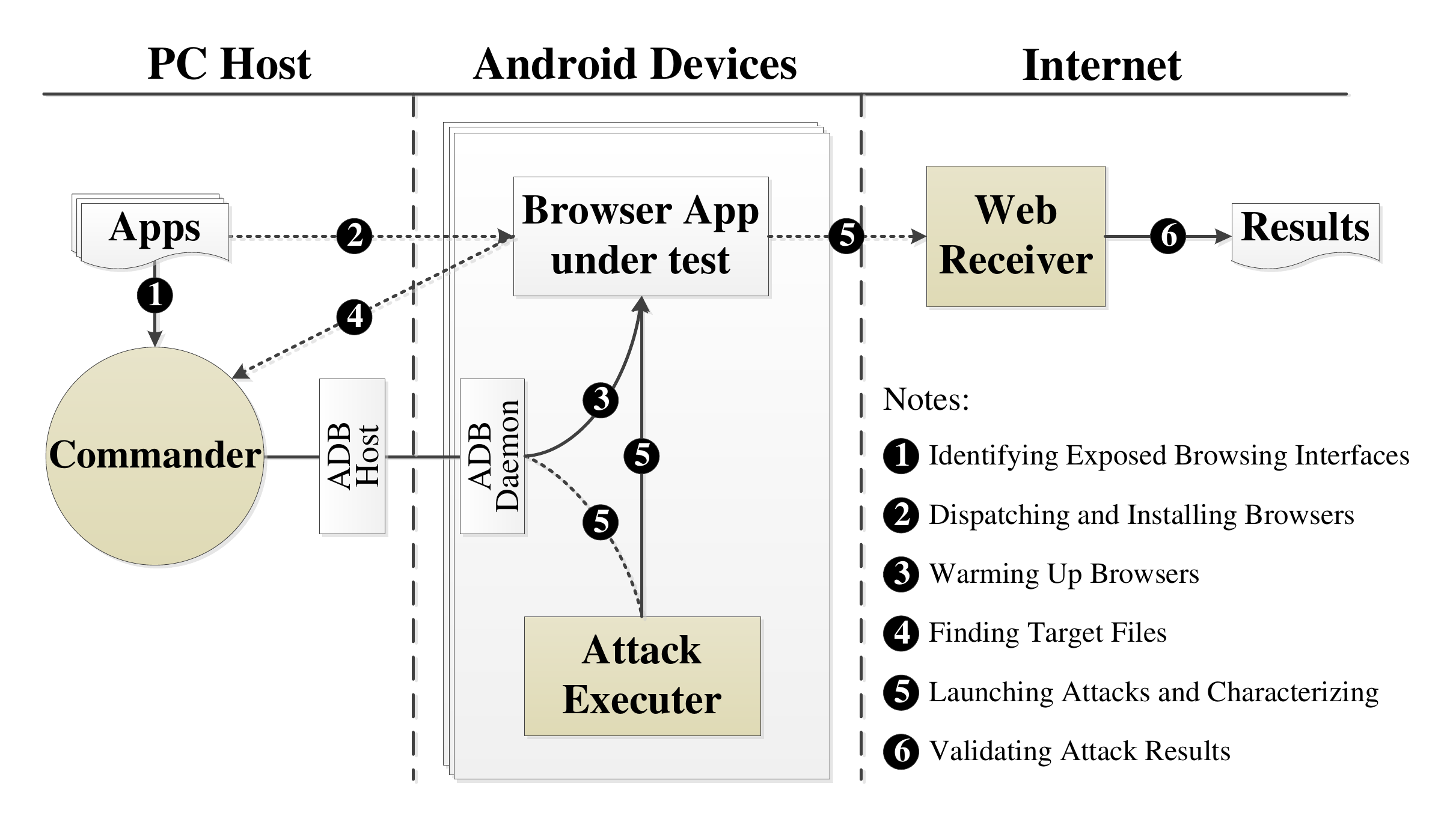}
\end{center}
\vspace{-6ex}
\caption{The architecture and workflow of our testing system.}
\vspace{-4ex}
\label{fig:system}
\end{figure*}

\myfig \ref{fig:system} shows the architecture and workflow of our testing system. The three main components in this system are \textit{Commander} for controlling the entire testing process, \textit{Attack Executer} for launching the \vul attacks, and \textit{Web Receiver} for validating whether the attacks are successful. The Commander running in a PC host controls the connected Android devices (which can be emulators or real phones) via Android Debug Bridge (ADB) channels (from ADB host to ADB daemons on devices). We implement Commander in the pure Python language to avoid the unstable issues of MonkeyRunner \cite{monkeyrunner} reported in \cite{SMVHunter14}. Moreover, we implement parts of the failure controlling mechanisms proposed in \cite{SMVHunter14} to improve the stability of ADB over long runtimes and use multiple threads to concurrently control each device for testing multiple Android versions in parallel.

We implement the Attack Executer as an Android app and install it in each tested device. Like a real attack app, it launches the \vul attacks to steal private files from the target browsers. Moreover, its attack behaviors are fully controlled by the Commander through each incoming attack command (including target browser information and attack parameters). Once receiving the attack commands, it generates the corresponding exploits on-the-fly and loads them into target browsers via the Intent channels. The Web Receiver, on the other hand, is a server-side program responsible for accepting the stolen private files and validating the attack results. An attack is considered successful if the stolen file is received.

%


\subsection{The Major Testing Steps}
\label{sec:workflow}

\myfig \ref{fig:system} shows six major testing steps in our system. We discuss them below in three pairs.

\noindent \textbf{Identifying exposed browsing interfaces}
We propose a lightweight but effective scoring mechanism to identify EBIs in Android browsers. The basic idea is to score each component based on our summarized EBI patterns in Section \ref{sec:atkcondition} and select the component with a maximal score as the EBI. That is, a component with the maximal score is most likely to be an EBI. This maximal score also helps us locate the major (or true) browsing interface. For instance, Chrome's \texttt{ManageBookmarkActivity} exhibits EBI patterns but is not functional for handling browsing requests. In this case, our scoring mechanism can help identify the right browsing interface \texttt{chrome.Main}, which shows more explicit EBI patterns, thus a higher score. When several EBIs have the same score, we handle such case by randomly selecting one EBI for dynamic testing. In addition, if all components score zero, we conclude there is no EBI in the browser. In our experiments, we find that this scoring mechanism can accurately identify the EBIs in 113 browsers out of the tested 116 browsers. For the remaining three cases that have no EBIs, one of them is only a browser add-on, and the other two do not expose their browsing interfaces.

\begin{figure*}[ht!]
\begin{center}
\vspace{-1ex}
\centerline{\includegraphics[width=1.1\textwidth]{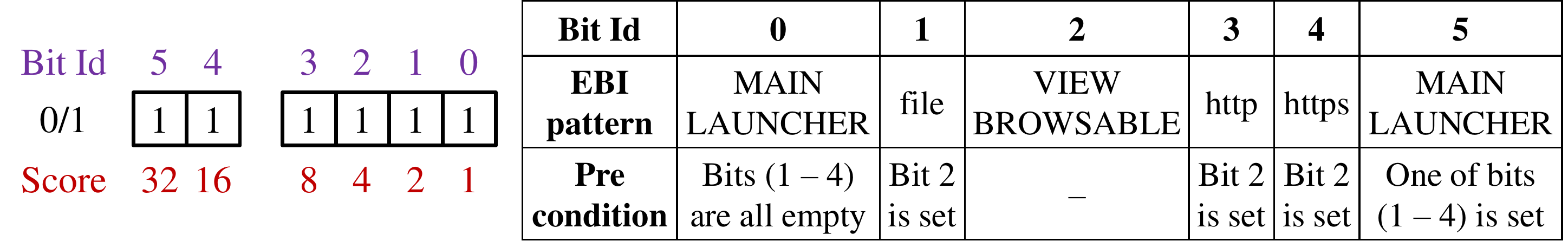}}
\end{center}
\vspace{-6ex}
\caption{The detailed rules for scoring EBI patterns using six bits.}
\vspace{-4ex}
\label{fig:EBIscore}
\end{figure*}

The detailed scoring algorithm works as follows. We use six bits to flag five specific EBI patterns (two bits are set for one pattern under different situations). \myfig \ref{fig:EBIscore} illustrates the detailed rules for scoring the EBI patterns under different scenarios. For example, if one component has an Intent filter which defines the \texttt{action} of ``VIEW'' and the \texttt{category} of ``BROWSABLE,'' we set bit 2 (i.e., a score of 4). If this Intent filter also registers the \texttt{data} scheme of ``http,'' we further set bit 3 (i.e., a score of 8). Now the component has a total score of 12, which can be used also for reversely inferring the EBI patterns using its binary representation.

These scoring rules (with different weights) are summarized according to our manual analysis of a dozen of EBI patterns.
First, we treat the basic EBI pattern (i.e., ``VIEW'' and ``BROWSABLE'') as a reference pattern.
On the basis of this pattern, we further assign weights to three data schemes, if any.
Among them, we score the ``https'' scheme higher than ``http,'' because we find accepting ``https'' is more likely to represent an EBI.  
On the other hand, we lower the ``file'' scheme even below the reference pattern, to remove the potential noises introduced by ``file''.
The noises can occur when ``file'' is registered for browsing document or video files.
So such components are actually document viewers or video players, instead of browser components.
Finally, we observe the ``LAUNCHER'' pattern, if exists, can add more weights when the aforementioned patterns also occur.
That is, a component with both ``BROWSABLE'' and ``LAUNCHER'' patterns will be always the major EBI, compared with those non-launcher ``BROWSABLE'' components. 
In addition, a component with only the ``LAUNCHER'' pattern should be scored less than other ``BROWSABLE'' components.

\noindent \textbf{Warming up browsers and finding target sensitive files}
The goal of warming up browsers is to produce some private files as the target sensitive files. To do so, the system automatically sends several normal browsing requests before launching the attacks. Specifically, the tested browsers are instructed to browse several Alexa top 10 websites using HTTP or HTTPS. This warming-up step can also help validate the EBIs identified by the scoring mechanism. That is, if an EBI is correctly identified, we can effectively warm up the corresponding browser. Otherwise, the browser will not respond according to our external browsing requests.

After warming up the browsers, our system continues to find as many target sensitive files as possible from the newly generated private files.  To do so, the system searches browsers' private file zones (i.e., \texttt{/data/data/package/}) using a set of prioritized keywords (e.g., ``cookie'', ``password'', and ``bookmark'') and certain file formats (e.g., ``.sqlite'' and ``.db'' files).
Note that accessing private file zones, which is normally disabled on unrooted phones, is only used for finding target sensitive files in our system (and attackers can also use this method to obtain the same information for their attacks). The actual \vul exploitation is still conducted by the Attack Executer through the normal Intent channels.

\noindent \textbf{Automatic attack validation and characterization}
Another challenge in designing our system is how to \textit{automatically} validate attack results and conduct further characterization. Unlike manual testing, we cannot rely on human intervention, such as naked-eye inspection. To address this issue, we pre-define patterns that describe the attack details given by the Commander and embed them into each attack request sent by exploit scripts, which will be finally received and interpreted by the Web Receiver. In particular, we embed five patterns into the attack requests: an app package's name (for identifying the tested browser), an attack ID (for differentiating different attacks), a device version (for characterizing attacks on different Android versions), contents (for transmitting and validating potential private files), and a key ID (for authentication and differentiating different experiments).


To further characterize the \vul attacks, we adopt the similar methods as for launching attacks, except that the attack scripts are now replaced by other scripts for characterization purposes. Specifically, we design HTML files to characterize the \file support (loaded from SD card or private file zones) and JavaScript execution in \file URLs. For example, the following HTML file is for characterizing the \file support. The Attack Executer loads this HTML file from both SD card and private file zones (with different attack IDs, such as \texttt{atk=5}), and sets the current Android version (e.g., \texttt{ver=4.3}).
\vspace{-1ex}
\noindent
\begin{verbatim}
<html><body>  <img src=`http://ourserver.com/req?pkg=example.package
&atk=5&con=reqflag&ver=4.3&kid=keyid'>  </body></html>
\end{verbatim}
\noindent
Interested readers may refer to Appendix \apJS~for the HTML file used to characterize JavaScript execution in \file URLs. It is relatively complex.

\section{Evaluation}
\label{sec:vulns}

\subsection{The Dataset and Experiments}

\noindent \textbf{Dataset}
Our dataset consists of 115 browser apps collected from Google Play on January 21, 2014.
Initially, we searched the keyword ``Browser'' on Google Play and fetched 139 browsers, after excluding several non-browser apps.
We further revisited these 139 browsers on March 21 to characterize their meta information (e.g., the install numbers) using the Selenium scripts \cite{Selenium}.
Based on the results, we further excluded 23 browsers in which
14 of them were no longer updated for more than one year,
and 9 others had been withdrawn from Google Play.
Among the remaining 116 browsers, one more was excluded, because it was only a browser add-on.

\noindent \textbf{Experiments}
We run our experiments using three Android phones:
Sony Xperia J (with Android 4.0), Google Nexus 4 (with Android 4.3), and Nexus 5 (with Android 4.4).
These phones are connected to a Dell Studio XPS desktop machine with Ubuntu 12.04 64-bit system through USB cables.
We do not use Android emulators in previous studies \cite{AppsPlayground13,NetworkProfiler13,SMVHunter14,ConColicAndroid12,Dynodroid13}, because they are not stable and a number of apps cannot be correctly installed or run on emulators.
However, accessing apps' private file zones via ADB on real phones is disabled by default. We thus root the phones to enable it for our automatic testing.

In this section, we report the results obtained from three independent experiment runs conducted on March 27 and June 18 (when the 4.4 device newly joined).
Our system incurs no false positives but may incur some false negatives due to the possible instability of dynamic testing.
To mitigate this possibility, our final result is a union of the results from these three runs.
Regarding the testing performance, each run takes around four hours (i.e., 3 minutes per app).
We use a relatively long timeout (12 seconds) before starting a new browsing request to obtain stable results and duplicate the app testing on three phones for observing possible different results in the three major Android versions.

\subsection{Vulnerability Results}

\begin{figure}[ht!]
  \vspace{-4ex}
  \centering
  \subfigure[The distribution of browsers with(out) vulnerabilities.] {
    \label{fig:total}
    \includegraphics[height = 35ex]{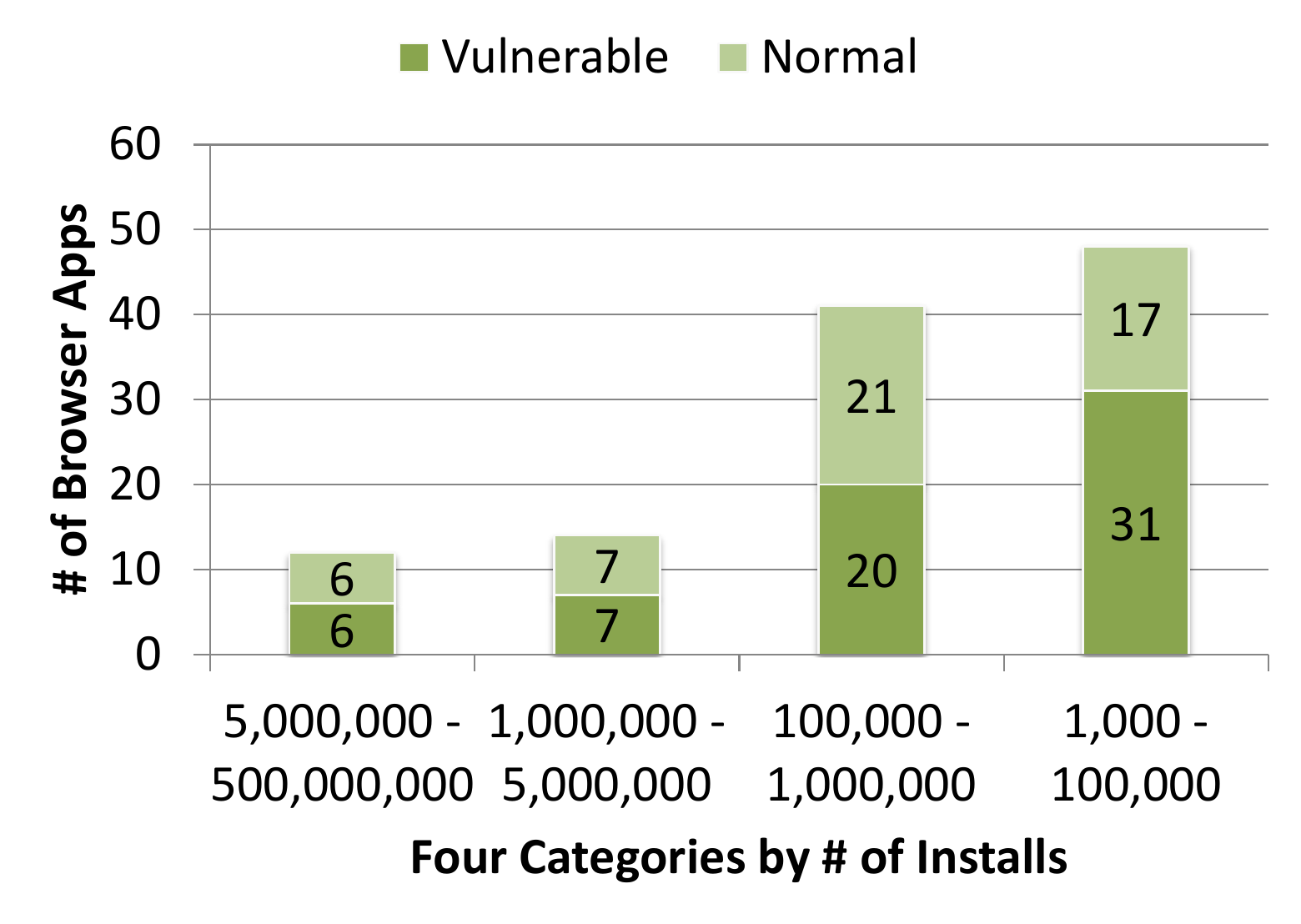}
  }
  \hspace{4ex}
  \subfigure[Detailed results for each attack.] {
	\label{fig:allvulns}
    \includegraphics[height = 35ex]{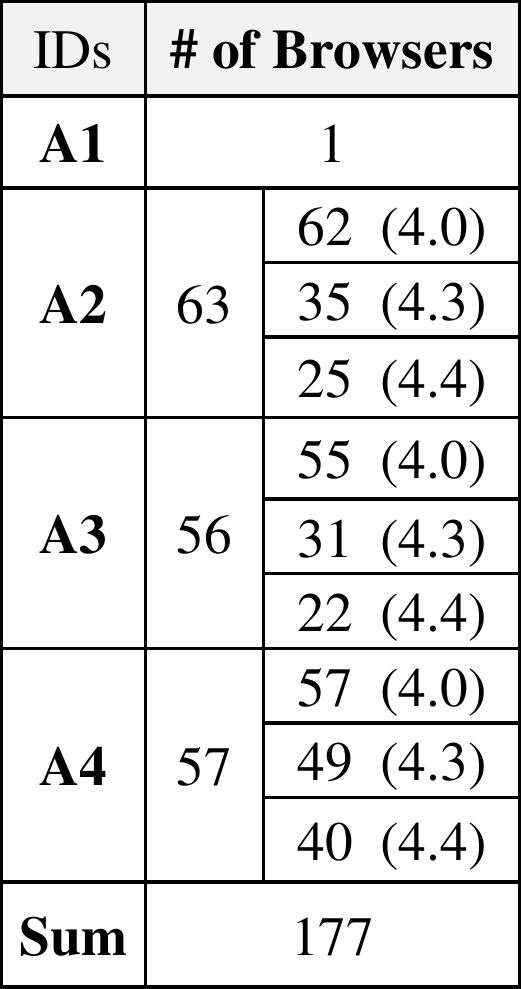}
  }
\vspace{-2ex}
\caption{Overall detection results in our dataset consisting of 115 Android browsers.}
\vspace{-4ex}
\label{fig:overall}
\end{figure}

\noindent \textbf{Overall results}
Our system identifies 64 vulnerable browsers and a total of 177 \vul issues, as shown in \myfig \ref{fig:total}.
The results clearly show that the vulnerabilities are prevalent in Android browsers: 55.7\% of browsers are affected and on average 2.77 issues per vulnerable browser.
Furthermore, according to their distribution by the number of installs, 13 out of 26 popular browsers with over million installs each are found vulnerable.
They are from top browser vendors, including Firefox, Baidu, and Maxthon. In other words, the \vul attacks are not easy to discover and were not known to them before our disclosures.

\myfig \ref{fig:allvulns} shows the detailed results for each \vul attack. 
In our dataset, we only discover one auto-file download issue, i.e., attack A1.
However, we observe that 71 browsers actually load and display the contents of their private files when challenged by attack A1.
Therefore, they will face the potential risk of screen-shot attacks, although we do not consider such risk as a vulnerability in this paper.
 
For attacks A2, A3 and A4, the number next to (4.0) (or (4.3) and (4.4)) is the number of browsers vulnerable to the attack on Android 4.0 (or 4.3 and 4.4). The number next to these three is the total number of vulnerable browsers for that attack. Some browsers are vulnerable on only one system. These three attacks have a similar number of vulnerable browsers, around 60.
Moreover, attack A4 is much less affected by different Android versions than A2 and A3. In the following sections, we thus do not differentiate the results of attack A4 on the three versions.
As for attacks A2 and A3, there are over 30 vulnerable browsers for each attack on Android 4.3 and over 20 on Android 4.4, mainly because the developers still use the old SDKs to compile their apps. Thus, their browsers cannot benefit from the webkit patch in Android SDK 4.1.


\noindent \textbf{Representative vulnerable browsers}
Table \ref{tab:majorvulns} summarizes 20 representative vulnerable Android browsers identified by our system.
To make it simple, we only use the app package name to refer to each browser, and their full app names can be obtained from Google Play.
We also include the number of installs for each browser to underscore the scope of the impact.
For each vulnerable browser, we list their detailed assessment results of the four \vul attacks launched by our system.
The red ``y'' means a successful attack, and the black ``n'', otherwise.
In addition, a blank space represents the case where our attack scripts cannot send response requests to our server, mainly because the target browser is either invulnerable or not stable on some Android versions (e.g., 4.3 and 4.4). For such cases, they are assumed invulnerable if no further manual efforts are involved.

\begin{table*}[ht!]
  \centering
  \vspace{-4ex}
  \caption{Representative vulnerable Android browser apps identified by our system.}
  \vspace{-2ex}
  \label{tab:majorvulns}
\scalebox{0.92}{
\begin{adjustbox}{center}
\begin{tabular}{ |c || c || c | ccc | ccc | c | c| }
\hline
\multirow{2}{*}{Categories} & \multirow{2}{*}{App Package Names} & \multirow{2}{*}{\ \ A1\ \ } & \multicolumn{3}{c|}{A2} & \multicolumn{3}{c|}{A3} & \multirow{2}{*}{\ \ A4\ \ } & \multirow{2}{*}{\# of Installs} \tabularnewline
\cline{4-9}
 & & & \ 4.0 & 4.3 & 4.4 \ & \ 4.0 & 4.3 & 4.4 \ & & \tabularnewline
\hline

\multirow{7}{*}{Popular} & org.mozilla.firefox      & \yes & & & & \no  & \no & \no   &  & 50,000,000 - 100,000,000 \tabularnewline
\cline{2-11}
 & com.baidu.browser.inter  & \no       & \yes &   & \no  & \yes &  \no & \no      & \yes  & 5,000,000 - 10,000,000 \tabularnewline
\cline{2-11}
 & com.mx.browser           & \no       & \yes & \yes & \yes   & \yes & \yes & \yes   & \yes  & 5,000,000 - 10,000,000 \tabularnewline
\cline{2-11}
 & com.jiubang.browser      & \no       & \yes & \yes & \yes   & \yes & \yes & \yes   & \yes  & 5,000,000 - 10,000,000 \tabularnewline
\cline{2-11}
 & com.tencent.ibibo.mtt    & \no       & \yes  &  &       & \no &   &     & \yes  & 1,000,000 - 5,000,000 \tabularnewline
\cline{2-11}
 & com.boatbrowser.free     & \no      & \yes & \yes & \yes & \no  & \no & \yes    & \yes  & 1,000,000 - 5,000,000 \tabularnewline
\cline{2-11}
 & com.ninesky.browser      & \no       & \yes & \yes & \yes & \yes & \yes & \yes   & \yes  & 1,000,000 - 5,000,000 \tabularnewline
\hline

\multirow{3}{*}{Tablet} & com.uc.browser.hd        & \no       & \yes & \yes & \yes   & \yes & \yes & \yes  & \yes  & 1,000,000 - 5,000,000 \tabularnewline
\cline{2-11}
 & com.baidu.browserhd.inter& \no       & \yes &    & \no & \yes &  \no & \no      & \yes  & 100,000 - 500,000 \tabularnewline
\cline{2-11}
 & com.boatbrowser.tablet   & \no       & \yes & \yes & \no  & \no & \no & \no & \yes  & 100,000 - 500,000 \tabularnewline
\hline

\multirow{3}{*}{Privacy} & com.app.downloadmanager  & \no       & \yes & \no & \no    & \yes & \no & \no    & \yes  & 10,000,000 - 50,000,000 \tabularnewline
\cline{2-11}
 & nu.tommie.inbrowser      & \no       & \yes & \yes & \yes   & \yes & \yes &   & \yes  & 500,000 - 1,000,000 \tabularnewline
\cline{2-11}
 & com.kiddoware.kidsafebrowser  & \no  & \yes & \no & \no    & \yes & \no & \no    & \yes  & 50,000 - 100,000 \tabularnewline
\hline

\multirow{3}{*}{Fast browsing} & \scalebox{0.9}{com.ww4GSpeedUpInternetBrowser} & \no       & \yes & \yes &     & \yes & \yes &     & \yes  & 1,000,000 - 5,000,000 \tabularnewline
\cline{2-11}
 & iron.web.jalepano.browser & \no      & \yes & \yes & \yes   & \yes & \yes & \yes    & \yes  & 500,000 - 1,000,000 \tabularnewline
\cline{2-11}
 & com.wSuperFast3GBrowser & \no      & \yes & \yes &     & \yes & \yes &    & \yes & 100,000 - 500,000 \tabularnewline
\hline

\multirow{4}{*}{Specialized} & com.appsverse.photon     & \no       & \yes & \yes & \yes   & \yes & \yes & \yes   & \yes  & 5,000,000 - 10,000,000 \tabularnewline
\cline{2-11}
 & com.isaacwaller.wikipedia  & \no       & \yes & \yes & \yes     & \no & \no & \no     &   & 1,000,000 - 5,000,000 \tabularnewline
\cline{2-11}
 & galaxy.browser.gb.free  & \no       & \yes & \yes &    & \yes & \yes &    & \yes & 100,000 - 500,000 \tabularnewline
\cline{2-11}
 & com.ilegendsoft.mercury  & \no       & \yes & \no & \no    & \yes & \no & \no    & \yes  & 100,000 - 500,000 \tabularnewline
\hline
\end{tabular}
\end{adjustbox}
}
  \vspace{-4ex}
\end{table*}

We organize these vulnerable browsers into five categories, mainly according to their popularity and unique features.
For example, in the ``Popular'' category, we present several popular browsers with over million installs each.
In particular, we identify an auto-file download issue (i.e., attack A1) in Firefox for Android, which is quite popular and has at least 50 million installs.
This security issue is ranked by Firefox a high impact one.
Moreover, we discover more \vul issues in other listed popular browsers.
For example, Maxthon Browser (\texttt{com.mx.browser}) and Next Browser (\texttt{com.jiubang.browser}) suffer from three \vul attacks in all Android versions we tested, which pose significant security threats to their five million users.

The second category (``Tablet'') lists three vulnerable browsers built for Android tablets.
Except for UC Browser HD (\texttt{com.uc.browser.hd}) that has over million installs, these browsers are not as popular as those in the ``Popular'' category.
However, we notice from Google Play that they are essentially the only choices for users who want to install a dedicated tablet browser.
This would entice attackers to launch more targeted attacks at tablet users.

Due to the page limit, the description on the last three categories of vulnerable browsers is available in Appendix \apNew.
Here we only mention two cases.
The Kids Safe Browser (\texttt{com.kiddoware.kidsafebrowser}) that provides children a safe Internet surfing environment by content filtering jeopardizes children's privacy by the \vul attacks.
Another example is a dedicated browser for browsing Wikipedia, called Wikidroid (\texttt{com.isaacwaller.wikipedia}).
Attackers can launch the \vul attacks to infer users' interests and profiles.

\subsection{Underlying Engine Analysis}

It is useful to determine how many browsers do not use the default engine (which has inherent flaws).
Implementing a custom web engine in Android usually requires embedding native codes as shared libraries (\texttt{.so} files).
For example, Chrome uses \texttt{libchromeview.so} as its underlying engine to support browsing functionalities.
Determining which \texttt{.so} files are web engines is hard and also beyond the scope of this paper.
Here, we adopt two strategies to infer which browsers embed their own engines.
First, we use regular expression ``\texttt{native.*loadUrl}'' to locate five browsers that implement their own native version of ``loadUrl'' API, including Chrome, Yandex (\texttt{libchromiumkit.so}), Flash Browser (\texttt{libxul.so}), and even the vulnerable UC Browser HD (\texttt{libWebCore\_UC.so}).
However, this strategy is not robust enough, because it even misses the Firefox engine.
Therefore, we directly inspect each \texttt{.so} file name from 24 browsers which have \texttt{.so} files.
The inspection (combined with existing knowledge) shows that another six browsers embed their own engines, such as Firefox (\texttt{libmozglue.so}), Dolphin Browser (\texttt{libdolphinwebcore.so}), and three Opera browsers (\texttt{libom.so}).

It is also a trend that more Android browsers will use custom engines.
Our analysis of five popular Chinese browser apps (which were collected on May 1) shows that four of them define their own engines.
They are QQ (\texttt{libmttwebcore.so}), Baidu (\texttt{libzeus.so}), Liebao (\texttt{libchromeview.z.so}) and Sogou (\texttt{libsogouwebcore.so}) browsers.
In particular, our system identifies Sogou Browser being vulnerable to \vul attack A4.

In summary, we have identified 15 (out of the total 120) browsers embedding their custom engines instead of the system default one.
In addition, our system identifies three of them being vulnerable: Firefox, UC Browser HD, and Sogou browsers.
These findings demonstrate the effectiveness of our system to uncover \file vulnerabilities in non-webkit browsers.

\section{Further Analysis and Recommendations}
\label{sec:characters}

\subsection{Analyzing the Patches}
\label{sec:patch}

\noindent \textbf{An overview}
We have devoted considerable efforts on reporting our identified vulnerabilities to the developers (see Appendix \apReport).
Table \ref{tab:patch} summarizes the nine patches received so far. Our analysis reveals three kinds of patch methods adopted by the developers.
First, similar to the method used by Chrome \cite{Symlink13}, Firefox's developer disabled the capability of accessing the contents of some unrenderable private files to address the auto-file download issue.
However, unlike Chrome, Firefox still allows \file access to the private file zone and loading renderable files.
We argue that accessing private file zone should be totally banned to mitigate all potential risks.
Second, Lightning Browser (\texttt{acr.browser.barebones}) and InBrowser (in its beta version, \texttt{nu.tommie.inbrowser.beta}) directly blocked the external \file URLs from other apps.
This fix suggests that supporting external \file URLs is not necessary for maintaining some browsers' functionalities.
It is interesting to note that the developer of Lightning Browser also applied this method to protect his two other browsers (one is a paid version, and the other an unpublished new browser).
Finally, the developers of most patched browsers chose to disable JavaScript execution in \file URLs, because it is the easiest way to thwart the three \vul attacks that require JavaScript support.
Although this patch does not eliminate all the possible risks (e.g., screen-shot attacks or origin-crossing attacks without JavaScript), it could be considered effective for the threat models considered in this paper.

\begin{table}[ht!]
  \centering
  \vspace{-4ex}
  \caption{An overview of the nine patches received from the developers.}
  \vspace{-2ex}
  \label{tab:patch}
\scalebox{0.935}{
\begin{adjustbox}{center}
\begin{tabular}{ |c | c | c| }
\hline
Package Names & Patched Versions & The Patching Methods \tabularnewline
\hline
org.mozilla.firefox & 28.0.1 & Disable accessing unrenderable private files \tabularnewline
\hline
acr.browser.barebones & 3.0.8a & Block external \file URLs and alert users \tabularnewline
\hline
nu.tommie.inbrowser.beta & 2.11-55 & Block external \file URLs \tabularnewline
\hline
com.baidu.browser.inter & 3.1.2.0 & Disable JavaScript execution in \file URLs \tabularnewline
\hline
com.jiubang.browser & 1.16 & Disable JavaScript execution in \file URLs \tabularnewline
\hline
com.baidu.browserhd.inter & 1.2.0.1 & Disable JavaScript execution in \file URLs \tabularnewline
\hline
easy.browser.classic & 1.3.6 & Disable JavaScript execution in \file URLs \tabularnewline
\hline
harley.browsers & 1.3.2 & Disable JavaScript execution in \file URLs \tabularnewline
\hline
com.kiddoware.kidsafebrowser & 1.0.4 & Disable JavaScript execution in \file URLs \tabularnewline
\hline
\end{tabular}
\end{adjustbox}
}
\vspace{-4ex}
\end{table}

\noindent \textbf{An interesting patching process}
During the process of analyzing the patches, we identified an interesting case which illustrates the importance of automatic testing even for patches.
The developers of Baidu Browser once sent us a version that they thought was patched, because they had disabled the JavaScript execution.
However, our system could still successfully exploit this ``patched'' version.
By a careful manual analysis of the patched version, we have found that there were two rendering points in Baidu Browser's browsing interface: one is invoked when users manually input a URL in the browser bar, and the other is for external browsing Intents.
Interestingly, the developers disabled the JavaScript support for \file URLs only for the first rendering point, thus leaving the real attack point intact.
Since the developers did not have an actual attack app, they tested the ``patch'' manually and mistakenly thought it was patched.

\subsection{Exposed Browsing Interfaces}

\begin{figure*}[ht!]
\begin{center}
\vspace{-4ex}
\includegraphics[width=0.6\textwidth]{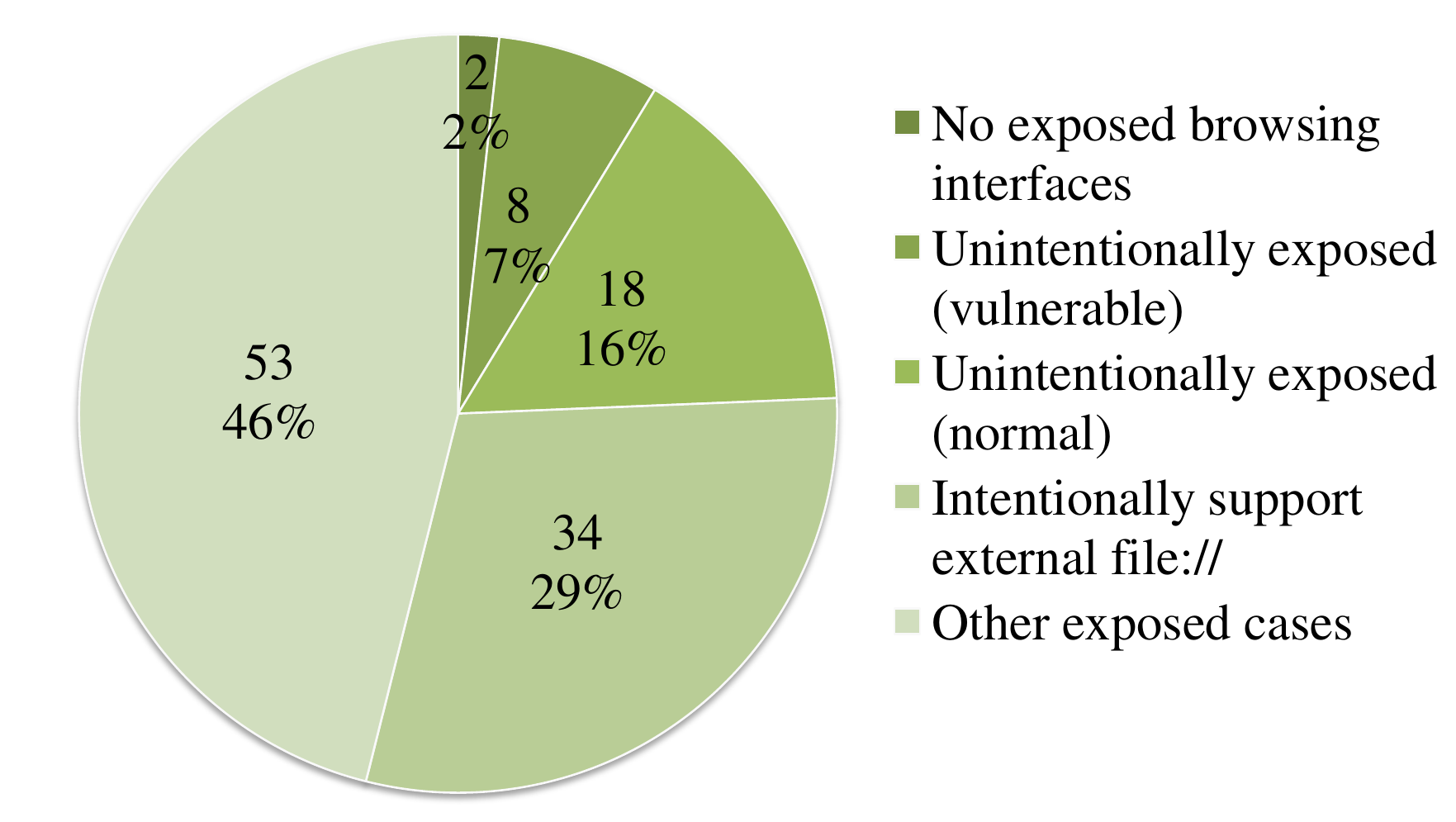}
\end{center}
\vspace{-6ex}
\caption{A breakdown of exposed browsing interfaces in the 115 tested browsers.}
\vspace{-4ex}
\label{fig:EBI}
\end{figure*}

\myfig \ref{fig:EBI} shows the breakdown of the EBIs in our tested 115 browsers, of which 113 expose their browsing interfaces, meaning that exposing browsing interfaces is a common practice among Android browsers.
However, we notice that 26 browsers (23\%) expose their browsing interfaces unintentionally.
Among them, eight are vulnerable.
In other words, these eight browsers could originally avoid the \vul issues, if they realized to close their unintentionally exposed interfaces.

We also observe that only 34 browsers (29\%) explicitly or intentionally accept external \file browsing requests.
But our dynamic testing actually finds 75 browsers supporting external \file browsing requests.
This discrepancy shows that the other 41 browsers may accidentally leak the \file channels to other apps. That is, they intend to support \file URLs only for internal uses (e.g., when users manually input a \file URL).

\subsection{\file Support in Android Browsers}

Based on our analysis, we report three major observations on the \file support in Android browsers.
First, (at most) 40 of our collected 115 browsers do not support \file at all.
It is worth noting that 40 is only a upper bound, because our system may not successfully characterize some browsers due to the limitation of dynamic analysis.
Among the 40 unsupported ones, Opera Mini and UC Browser Mini are the very popular ones.
Opera Mini explicitly mentions ``\emph{The protocol ``file'' is not currently supported}'' when a \file URL is entered, whereas UC Browser Mini redirects users to a Google search page using the keyword of the entered URL.
Other unsupported cases that we manually confirm are dedicated browsers, such as The Pirate Bay Browser for browsing torrents and SkyDrive Browser for accessing Microsoft's SkyDrive service. These cases collectively show that \file is generally not supported in lightweight and dedicated browsers, and this practice spares them from the \vul attacks.

Second, we find that several popular browsers already forbid \file access to private file zones.
Our system identifies four such cases, including Chrome, Dolphin (\texttt{mobi.mgeek.TunnyBrowser}), UC (\texttt{com.UCMobile.intl}) and Yandex browsers.
All of them allow \file access to contents in SD card and permit JavaScript execution in \file URLs, but forbid \file access to their private file zones.
Thanks to this security policy, they are robust to most \vul attacks (i.e., except A2).
We therefore recommend adopting this practice for all Android browsers, because it can better meet the security model of mobile systems.

Finally, we observe three browsers actively disabling the JavaScript execution in \file URLs: 3G Browser (\texttt{com.mx.browser.free.mx100000004981}) and another two from the same developer (Maxthon Tablet and Maxthon Fast Pioneer browsers). Although the percentage of this practice is currently low (i.e., 3 out of 75), according to our analysis of the patches, we believe that more browsers will follow this practice.

\section{Related Work}
\label{sec:related}

\noindent \textbf{WebView security}
The closest related works are those on the security of WebView, which uses Android's default web engine (mainly webkit) APIs to help apps display web pages.
However, different from our study, most of these studies (e.g., \cite{WebViewAtk11,Bifocals13,NoFrak14}) mainly concern the insecure invocation between JavaScript and Java levels which may compromise a WebView app by misusing its exposed JavaScript interfaces.
In particular, the file-based cross-zone scripting attack reported in \cite{Bifocals13} is similar to the \vul attacks, but their attack follows the man-in-the-middle model where malicious JavaScript codes are injected by network adversaries.
Without adopting a realistic threat model and proposing detailed attacks, they conclude that file-based cross-zone scripting vulnerabilities are \textit{fortunately} fairly rare.
In our study, however, we show that \file vulnerabilities are prevalent in Android browsers.
Additionally, our study is more general for testing major practices in the Android browser ecosystem (i.e., not limited to WebView flaws), and we also identify non-webkit vulnerable cases (notably Firefox and UC Browser HD).


\noindent \textbf{Android exposed component issues}
One important condition for launching \vul attacks is that browsing interfaces in victim browsers are exposed.
Many previous works (e.g., \cite{ComDroid11,Woodpecker12,CHEX12,ContentScope13,ICCEpicc13,CustomRom13}) have studied the general exposed component problem from the perspective of information flow analysis.
They aim at the source-sink problem that other apps can trigger dangerous APIs (i.e., sinks) in an exposed component from its exposed entry points (i.e., sources).
Compared to the \vul attacks, constructing their exploits are less complicated (due to the main focus on the raw Intent fields) and do not require the domain knowledge of browser SOP and file protocol.
The exploit for Facebook Next Intent issue in \cite{CrossOrigin13} is also launched from \file, but it does not aim at stealing Facebook app's private files as the Facebook \vul attack reported in \cite{Facebook13}.

\noindent \textbf{Android dynamic testing}
Besides our system, there are a number of other Android dynamic testing systems proposed for various purposes.
Systems from the software engineering community aim at improving the app test rates by covering more code paths (e.g., \cite{ConColicAndroid12,Dynodroid13,ASE13}) with lower costs \cite{SwiftHand13} and in more flexible ways \cite{PUMA14}.
In contrast, systems for security testing focus on adding more dedicated components, such as taint tracking in \cite{AppsPlayground13}, fingerprint generator in \cite{NetworkProfiler13}, and pre-performed static analysis in \cite{SMVHunter14}.
In our case, we also embed an EBI scoring module and two dedicated components (i.e., the Attack Executer and Web Receiver) into our system, making it the first system for detecting the \file vulnerabilities in Android browsers.

\section{Conclusions and Future Works}
\label{sec:conclude}

In this paper, we identified a class of attacks in Android called \vul that exploits the vulnerable \file to obtain user's private files, such as cookies, bookmarks, and browsing histories. We designed and implemented an automatic system to detect the vulnerabilities in 115 browser apps.
Our results show that the vulnerabilities are prevalent in Android browsers. More than half of our tested 115 browser apps were found vulnerable.
A further detailed analysis yielded more insights into the current browser practices, such as exposed browsing interfaces and allowing \file access to private file zones. Our vulnerability  reports also helped around ten developers patch their vulnerable browsers promptly. For one browser, our system helped discover that their first patch failed to block the vulnerability.

Our system currently focuses on detecting \file vulnerabilities in Android browsers. However, the \vul attacks may also exist in other kinds of apps that use web engine APIs. For example, Facebook was identified vulnerable to attack A2 \cite{Facebook13}, although it only suffered with another issue called Next Intent \cite{CrossOrigin13}.
Detecting \file vulnerabilities in these non-browser apps is a future work of our system.
We plan to incorporate static analysis techniques to identify ``similar'' browsing interfaces which may not have clear EBI patterns.

There are another two limitations in our current system and experiments.
First, some browsers have the splash or welcome views in the front of their browsing interfaces, which may interfere with our automatic attacks.
But we also notice several such cases (e.g., Next and Boat browsers) that actually do not affect the effectiveness of our attacks, because the underlying component is still the browsing interface although it is not visible.
Second, our current experiments do not cover the default browsers which are pre-installed in devices, because we do not have enough phones to collect and test them.

\bigskip
\noindent \textbf{Acknowledgements}
We thank the three anonymous reviewers for their critical comments.
This work is partially supported by a grant (ref. no. ITS/073/12) from the Innovation Technology Fund in Hong Kong.

\begin{footnotesize}
\bibliographystyle{splncsnat}
\bibliography{esorics}
\end{footnotesize}

\section*{Appendix}

%

\subsection*{\apJS \hspace{4mm}Another Example of Characterizing Scripts}
The HTML file for characterizing JavaScript execution in \file URLs is given below.
\vspace{-1ex}
{\small
\noindent
\begin{verbatim}
<html><body>
<script>
var d = document; var img = d.createElement(`img');
img.src = `http://ourserver.com/req?pkg=example.package&atk=7&con=reqflag
&ver=4.3&kid=keyid';
d.body.appendChild(img);
</script>
<noscript>
<img src=`http://ourserver.com/req?pkg=example.package&atk=7&con=
&ver=4.3&kid=keyid'>
</noscript>
</body></html>
\end{verbatim}
\noindent
}

\vspace{-4ex}

\subsection*{\apNew \hspace{4mm}The Description on The Other Categories of Vulnerable Browsers}

In the third category (``Privacy'') of Table \ref{tab:majorvulns}, three representative vulnerable browsers built for users' privacy are selected.
For example, Downloader \& Private Browser (\texttt{com.app.downloadmanager}) is a quite popular browser that provides password protection for users' private files.
However, other app can access and steal these sensitive files by exploiting this app, contradicting their users' original intention of using this browser to protect their privacy.
Similarly, InBrowser (\texttt{nu.tommie.inbrowser}) is a browser supports incognito browsing by default. 
As another important example, the Kids Safe Browser (\texttt{com.kiddoware.kidsafebrowser}) that provides children a safe Internet surfing environment by content filtering jeopardizes children's privacy by the \vul attacks.

Some users prefer the browsers that are optimized to provide fast browsing experience.
We summarize three such vulnerable browsers in the category of ``Fast browsing''.
4G Speed Up Browser (\texttt{com.ww4GSpeedUpInternetBrowser}) and 3G Speed Up Browser (\texttt{com.wSuperFast3GBrowser}) are vulnerable to all attacks A2, A3, and A4.

In the last category of ``Specialized,'' Photon Flash Player \& Browser (\texttt{com.appsverse.photon}) and Galaxy Flash Browser (\texttt{galaxy.browser.gb.free}) are quite popular due to their dedicated support of Flash player in Android browsers.
However, both of them are vulnerable to three kinds of \vul attacks in all Android versions (except the latter cannot run stably in 4.4).
The second case is a dedicated browser for browsing Wikipedia, called Wikidroid (\texttt{com.isaacwaller.wikipedia}), allowing users to save their browsing bookmarks.
Attackers can launch the \vul attacks to steal these bookmarks and use them to infer users' interests and profiles.
The last case is called Mercury Browser (\texttt{com.ilegendsoft.mercury}) which is selected for its popularity in
its iOS version. We suspect that some Android users who have migrated from iOS system will possibly install this browser.

\subsection*{\apReport \hspace{4mm}Vulnerability Reporting}
\noindent \textbf{Our reporting process}
As mentioned in Section \ref{sec:patch}, we have spent considerable efforts on reporting our identified vulnerabilities to the developers.
Our reporting process was started on February 7 and is still ongoing at the time of writing.
So far we have reported 39 vulnerable browsers, including all the representative ones in Table \ref{tab:majorvulns}.
We continue to report the remaining cases and will release a project website to help developers better understand the vulnerabilities.

We report each case mainly in three steps.
First, we send a notice email to the email address recorded in Google Play, mentioning that we have identified vulnerabilities in their browsers without details.
After receiving their replies and confirming their identities, we explain the \vul attacks that their browsers are vulnerable to.
Finally, if they send us a patch, we analyze it using our system and report to them again whether their patch can correctly block the vulnerability. For the unresponsive developers, we contact them again until receiving their responses.

\noindent \textbf{Responses from developers}
We have currently received 19 replies from the developers.
Among them, 15 gave us further responses after being notified of the vulnerability details, and all of them confirmed our vulnerability reports.
In particular, we have received two bug bounty gifts from Baidu company.
Some excerpts of developers' responses can be found in Appendix \apRes.
Regarding the aforementioned Firefox issue for attack A1, our discovery of this serious vulnerability was also made independently by another security expert. We were told by Mozilla that ``\emph{this flaw has been fixed in the latest version of Firefox for Android, version 28.0.1}'' just the day before our reporting \cite{Firefox14}.

\subsection*{\apRes \hspace{4mm}Excerpts of Developers' Responses}

Besides acknowledging our reports, developers of Baidu Browser are also interested in our automated testing system, and their feedbacks are as follows.
\begin{quote}
``\emph{Thank you for your reporting. We confirm that those three vulnerabilities affect Baidu Browser inter version, but do not affect its Chinese version. Please provide us your contact address so we can send a gift for your nice work.}''\\
``\emph{I am security architect in Baidu Mobile-App-Team. your work is really valuable for us. further, please also scan our Baidu relative apps, \ldots}'' \\
``\emph{Just as you say, the tablet version also suffers this vulnerability. We will fix it soon, and give you the patched version. \ldots We will send a gift to you for your excellent work.}''
\hfill -- Responses from Baidu Browser
\end{quote}

Some developers keep us updated about their process working on the patches, such as the vendors of InBrowser and Kids Safe Browser.
\begin{quote}
``\emph{We're very grateful for the detailed error-report and our engineers are working on the issue as we speak. We'll publish those changes silently in our Beta stream to start with and then publish publicly within a couple of weeks.}''
\hfill -- Responses from InBrowser
\end{quote}

Moreover, we notice some individual developers are more responsible for their security issues than some big companies.
For example, the student developer of Lightning Browser and the individual developer of Easy Browser always notify us of their patching updates.
\begin{quote}
``\emph{I'm a one man development team, so I handle everything. I'm really interested in the details of these vulnerabilities. P.S. Unfortunately, I can't offer any monetary compensation for discovering the vulnerabilities since I'm just a poor student and this browser is a side project.}'' \\
``\emph{Ah, thanks for the clarification, Daoyuan. I'll see about what I can do.}'' \\
``\emph{I have modified it to block all external requests to load file urls, which should block the vulnerabilities.}''
\hfill -- Responses from Lightning Browser
\end{quote}


\end{document}